\documentstyle[12pt,epsf]{article}
\newcommand{\be}{\begin{eqnarray}}
\newcommand{\ee}{\end{eqnarray}}
\newcommand{\nn}{\nonumber}

{
{

\def\0n{0\nu\beta\beta}

\def\l{M_{L\hskip-1.5mm/}}

\begin{document}
\thispagestyle{empty}
\begin{center}
{\Large \bf Majorana neutrinos with split fermions
in extra dimensions} \\
\vspace{1.5cm}
{\large H.V. Klapdor-Kleingrothaus $^{\dagger}$
and U. Sarkar$^{\dagger,\ddagger}$}\\
\vspace{0.8cm}
{$^{\dagger}$ Max-Planck-Institut f\"ur Kernphysik,
P.O. 10 39 80, D-69029 Heidelberg, Germany} \\ \vspace{0.3cm}
{$^{\ddagger}$ Physics Department, Visva-Bharati University,
Santiniketan 731 235, India}
\end{center}

\vspace{2.5cm}

\begin{abstract}

We propose new solutions to the neutrino mass problem in 
theories with large extra dimensions in a thick wall scenario. 
It has recently been
argued that our 3-brane could be a thick wall at the 
boundary of the bulk. The gauge bosons and the Higgs
scalars have an almost flat profile on this wall, while
fermions could have localized profile with left-handed
and right-handed components displaced with respect to each
other. We point out that
with split fermions it is possible to generate
Majorana neutrino masses contributing to the neutrinoless 
double beta decay. The almost degenerate neutrinos
can also come out naturally in this case.
Unlike other models of neutrino masses in extra dimensions
there are no bulk fields in this scenario. 

\end{abstract}

\vspace{1cm}
\newpage
The recent developments in theories with extra dimensions
have changed our conventional idea about physics beyond the
standard model \cite{ed1}. In usual theories the weakness of gravity
is attributed to the very small coupling of the gravitational
interaction. This leads to the very high Planck scale where gravity could
become strong and could then influence the standard model.
In theories with large extra dimensions one assumes that
there are extra dimensions in which gravity propagates, 
whereas the ordinary standard model particles live in a
4-dimensional wall at the boundary of the extra dimensions.
Then the gravitational interaction strength in the bulk
could be strong and the overlap of the graviton wave function
in our 3-brane at the boundary makes gravity weak in our world.
This would then allow a very low fundamental scale of about
TeV replacing the effective Planck scale in the theory. 

The main advantage of these theories with extra dimensions
and TeV scale gravity is that there is no gauge hierarchy 
problem, but there is new problem of absence of any
large scale. The smallness of neutrino mass is usually
attributed to the large lepton number violating scale. But
there is no large scale in the theories with extra 
dimensions, so these mechanisms 
can not be used. Consider the effective 4-dimensional
operator in the standard model, which gives mass to the
neutrinos \cite{op}
\begin{equation}
{\cal O}_{eff} = {f_{ij} \over \l} \ell_{iL} \ell_{jL}
\phi \phi . \label{op1}
\end{equation}
Since the highest scale in the theory of extra dimensions
is the fundamental 
scale $M_*$, which is of the order of TeV, the effective
neutrino mass comes out to be fairly large now, unless
we can make the effective coupling constant $f_{ij}$ to
be small.  

There are several solutions to this problem, proposed with
the same philosophy to make the coupling small \cite{nu1,nu2,nu3}. 
In theories with extra dimensions all standard model particles 
reside in the four-dimensional wall. Gravity now propagates in
the higher-dimensional space and the overlap of the wave function
of the gravitons with the four-dimensional wall is very small. So,
the gravity coupling to matter in our world is suppressed 
by the volume of the extra dimensions. Similar to 
gravity if there are some bulk particles which move in all 
dimensions (which constitute the bulk of space in the
extra dimensions as compared to our wall which is confined 
only at one end), their overlap in
our brane would be small and that can give a small neutrino
mass in our brane \cite{nu1,nu2}. This gives a Dirac mass to the
neutrinos. In another scenario lepton number is
broken in another distant brane, or in the bulk,
which is then conveyed to
our brane by a bulk scalar field \cite{nu1,nu3}. The profile of the 
bulk scalar then can give a small effective coupling constant.

Recently it has been suggested that to solve the 
fermion mass hierarchy one could consider a thick
wall scenario, in which the left and right-handed components
are localized at different points with small overlap 
in the thick wall \cite{thw1,thw2,thw3,thnu}. 
Our 3-brane now has a spread (unlike other models where our
3-brane is confined to one point in the extra dimensions) in the
extra dimensions. The gauge and the Higgs bosons 
can propagate anywhere within this thick brane and
they have an almost constant profile in our brane.
Only outside this thick wall their profile falls 
off exponentially. However, within this thick wall 
the fermions are confined at different points with definite
profiles. The overlap of the different fields in any interaction
then gives the hierarchical Yukawa couplings \cite{thw1,thw2}.

We extend this scenario to explain the smallness of
the neutrino mass and show that tiny Majorana neutrino
masses come out naturally from this scenario. We do not
require any bulk fields to make the neutrino mass small.
In the present scenario
we break lepton number in our brane at the fundamental scale, 
but because of the small overlap of wave functions of the
required fields with each other 
the effective coupling constant $f_{ij}$ 
comes out to be very small naturally. 

The thick wall scenario was proposed to solve the problem of 
hierarchy of fermion masses \cite{thw1}. 
The fermions are localized at different
points in the higher-dimensional space. This can come from
string theory depending on the construction of the p-branes,
but there is also a field-theoretical realization of this idea.
Consider a five-dimensional example, with 
$z = \{x,y\}$ and $y$ as the coordinate of the fifth dimension,
in which a five-dimensional fermion $\Psi $ and a scalar 
${\cal S}$ couple through a Yukawa coupling term
$\sim \int d^5 z {\cal S} \bar \Psi \Psi$. If the 
scalar field has a position-dependent vacuum expectation value,
which changes sign at a point $y_0$ in the extra dimension,
then the fermion will be localized at $y_0$ with a Gaussian
profile in the extra dimension centered around $y_0$
\begin{equation} 
\Psi (x,y) = A~ e^{- \mu^2 (y - y_0)^2 } ~ \psi(x) ,
\end{equation}
where $\psi$ is a normalized four-dimensional massless
left-handed fermion, $A = (2 \mu^2 / \pi)^{1/4}$ is the normalization
and $\mu = \sqrt{\partial <{\cal S} >/2}$ is related to the slope of
the scalar field profile. $y_0 =0$ for a massless five-dimensional
field $\Psi$, but when a mass term is added for a particular
fermion field $\int d^5 z M_i \bar \Psi_i \Psi_i$, that field is localized 
at $y_0 = y_i = M_i/2 \mu^2$. 

A generic five-dimensional Yukawa interaction now gives
\begin{equation} 
{\cal L}_Y = \int d^5 z \sqrt{L} \kappa \Phi \bar \Psi_i \Psi_j
= \int d^4 x \lambda \bar \psi_i \psi_j \phi ,
\end{equation}
where $L$ is the domain wall width and the effective four-dimensional
Yukawa coupling constant comes out to be
\begin{equation} 
\lambda = \int dy \kappa A e^{- \mu^2 (y - y_i)^2 }
A e^{- \mu^2 (y - y_j)^2 } = \kappa e^{- \mu^2 (y_i - y_j)^2/2} .
\end{equation}
In general, $\kappa$ could depend on the indices $i,j$, but for
purpose of simplification
it is assumed that there is only one constant. $\phi$ is the
standard model Higgs doublet contained in the five-dimensional
scalar $\Phi$. The Gaussian width $\mu^{-1}$ has to be much larger
than the wall thickness $L$ for this mechanism to work, but for the
field theoretic description to work there is a limit $\mu^2 L < M_*$,
where $M_*$ is the fundamental scale in the problem. Combining with
other constraints, the requirement that the 
Yukawa coupling to be perturbative at $M_*$ now gives
\begin{equation} 
\mu < M_* < 1000 L^{-1} ~~~~~{\rm and}~~~~~ L^{-1} < \mu < 30  L^{-1} .
\end{equation}
Constraints from flavor changing neutral currents mediated by
the Kaluza-Klein gauge bosons constrain the wall thickness
$~~L^{-1} \geq 100$ TeV. 

Let us now consider the neutrino sector. Although the neutrino 
masses were discussed in the context of thick wall scenarios
\cite{thnu}, our proposed mechanisms differ from them. Here we
do not require any bulk particles and the neutrinos get
a lepton number violating Majorana mass. Moreover,
in this scenario the neutrino masses could be almost degenerate,
so that they can contribute to the dark matter of the universe
and also explain the neutrinoless double beta decay \cite{ndb,ndban}.
In other models of neutrino masses in extra dimensions there
is no natural mechanism to explain the almost degenerate neutrinos.
Even in ordinary theories it is difficult to accomodate an 
almost degenerate Majorana neutrino naturally. Starting from
a grand unified theory and if one evolves the Yukawa couplings
in supersymmetric models, it becomes difficult to 
maintain the degeneracy \cite{lola}. First we propose two 
different possibilities, each of which has some difficulties.
Then we consider a more general model combining both the
mechanisms, which has several interesting features.

First we introduce only
right-handed neutrinos for three generations $N_{\alpha R}$, 
$\alpha = 1,2,3$, which
are singlets under the standard model gauge group. We
then allow all possible renormalizable interactions consistent with
the standard model gauge symmetry. The Majorana mass term of the 
right-handed neutrinos will then violate lepton number and set the scale
of lepton number violation. The interactions of the right-handed 
neutrinos are given by
\begin{equation}
{\cal L}_N = M_{N \alpha \beta} N_{\alpha R} N_{\beta R} + 
h_{i \alpha} \bar l_{iL} N_{\alpha R} \phi ,
\end{equation}
where $l_{iL}$ is the left-handed lepton doublet and the Majorana 
mass of the right-handed neutrinos violate lepton number $M_N = \l$.
We have written this interaction in terms of
four-dimensional fields. The fifth dimension has been integrated
out to get these effective coupling constants $h_{i \alpha}$ and
the Majorana mass term $M_{N \alpha \beta}$. At this stage, this
is exactly similar to the usual see-saw mechanism of neutrino 
masses \cite{seesaw}.

Since any lepton number violating effective interactions of the 
left-handed scalars can originate only from these two interactions,
the lepton number violating mass scale ($\l$) in equation (\ref{op1})
should be given by $M_N$ and the coefficients $f_{ij}$ should be
determined by $h_{i \alpha}$. So, although the Majorana mass term
of $N_R$ would allow the effective lepton number violating operator 
(\ref{op1}) with very little suppression from the scale of lepton
number violation $\l < M_*$, the effective coupling now could be
very small. 

The above mentioned effective four-dimensional 
interactions come from five-dimensional terms in the Lagrangian
\begin{equation}
{\cal L}_{Yuk} = \int d^5 z ~ \left[ M_5 \Psi_{N \alpha}
\Psi_{N \beta} + \sqrt{L} ~ \kappa ~ 
\bar \Psi_{li} \Psi_{N \alpha} \Phi \right] ,
\end{equation}
so that 
\be
M_{N \alpha \beta} &=& M_5 ~ {\rm exp} \left[- \mu^2 {(y_{N \alpha} -
  y_{N \beta})^2 \over 2} \right] , \nn \\
h_{i \alpha} &=&  \kappa ~ {\rm exp} \left[- \mu^2 {(y_{li} -
  y_{N\alpha})^2 \over 2} \right] .
\ee
The diagonal elements of $M_N$ are all the same and equal to $M_5$. 
The right-handed neutrinos could be separated in space so
that the mass matrix is diagonal and given by an identity matrix.
But in general they could be neighbours and the mass degeneracy 
could be broken. We demonstrate this with an example. 

Consider the charged lepton mass matrix to be diagonal. Then
the required mass hierarchy could be achieved with the configuration
of the left-handed lepton doublets $l_i$ and the right-handed charged 
leptons $e_i$ given by (similar to ref. \cite{thw1})
\begin{equation}
l_i = \mu^{-1} \pmatrix{ 12 \cr 0 \cr -1 } ;
~~~~ {\rm and} ~~~~ e_i = \mu^{-1} \pmatrix{ 6.87 \cr
3.95 \cr -4.15 } .
\end{equation}
We further assume that the multiplets of the same $SU(2)_L$ 
representations to be located at the same place
in the fifth dimension. Then both
$e^-_{iL}$ and $\nu_{iL}$ (contained in $l_i$) 
will be located at the same place
with the same profile, and the configurations of $\nu_{iL}$ are 
the same as given for $l_i$. If the
right-handed neutrinos are now localized at
$$ N_\alpha = \mu^{-1} \pmatrix{ 7.2 \cr 5 \cr 4.8 } , $$
the Yukawa couplings and the right-handed Majorana mass
matrix will be given by
$$ h_{i \alpha} = \pmatrix{0.093 & 5.7 \times 10^{-7} & 1.3 \times
10^{-7} \cr 5.7 \times 10^{-7} & 0.093& 0.248 \cr 3.2 \times 10^{-10}
& 3.8 \times 10^{-4} & 0.0012 } $$ and $$ M_{N \alpha \beta} = M_0
\pmatrix{ 1& 0.135 & 0.0059 \cr 0.135 & 1 & 0.486 \cr 0.0059 & 
0.486 & 1 }  ,$$ where $M_0 = \l < M_*$ is the scale of lepton number
violation. In this example we considered $M_0 \sim 10^6$ GeV.
The eigenvalues of the left-handed neutrino mass matrix
\begin{equation}
m^\nu_{ij} =  {h_{i \alpha} h_{j \beta}^T <\phi>^2 
\over M_{N \alpha \beta} }
\end{equation}
come out to be $10^{-9}$ eV, $0.0062$ eV and $0.065$ eV, which
are the masses required to explain the atmospheric neutrino 
anomaly and the large mixing angle solution of the solar neutrinos. 
The correct mixing angles come out only when the charged
lepton mass matrix is not diagonal or 
the neutrinos are spread over more than one extra dimension. 

We now consider another possibility of Majorana neutrino mass
generation in the thick wall scenario. Instead of a right-handed
neutrino we now introduce a triplet Higgs scalar in the theory
\cite{trip}.
The interactions of the five-dimensional triplet Higgs scalar
($\Xi (z)$) now violate lepton number explicitly 
\begin{equation}
{\cal L}_\Xi = \int d^5 z \left[ \sqrt{L} ~ \kappa_l \Xi 
\Psi_{l i} \Psi_{lj} + \mu_\Xi \Xi^\dagger \Phi \Phi + M_\Xi 
\Xi^\dagger \Xi \right]  .
\end{equation}
From these interactions we can derive the four-dimensional 
effective interactions
\begin{equation}
{\cal L}_\xi = \int d^4 x \left[ f_{ij} \xi 
\psi_{l i} \psi_{lj} + \mu_\xi \xi^\dagger \phi \phi 
+ M_\xi \xi^\dagger \xi \right]  .
\end{equation}
This gives a Majorana neutrino mass to the left-handed
neutrinos
\begin{equation}
m^\nu_{ ij} = - f_{ij} \mu_\xi {< \phi>^2 \over M_\xi^2} .
\end{equation}
The effective four-dimensional Yukawa coupling constant 
$f_{ij}$ is given by
\be 
f_{ij} &=& \int dy \kappa_l~ e^{- \mu^2 (y - y_\xi)^2 } ~
e^{- \mu^2 (y - y_i)^2  }~ e^{- \mu^2 (y - y_j)^2  }  \nn \\
&=&
\sqrt{3 \over 2} ~\kappa_l ~ {\rm exp} \left[ - {\mu^2 \over 2}
(y_i - y_j)^2 
\right] 
\cdot {\rm exp} \left[ - {2 \over 3} \mu^2 
\left( y_0 - {y_i + y_j \over 2} \right)^2 \right]  . 
\ee
$y_i$ and $y_j$ are the positions at which the leptons are 
localized and $y_0$ is the point in the fifth dimension
where the triplet Higgs is localized. The first term gives 
a suppression depending on the separation of the two neutrinos,
while the second term gives a suppression depending on the
average separation of the neutrinos compared to the triplet
scalar. For the diagonal elements, the first term is identity,
while for the off-diagonal elements they are almost zero. So,
in the basis in which the charged lepton mass matrix is 
diagonal, all the off-diagonal elements vanish and the 
neutrino mass matrix always comes out to be diagonal. 
Since the standard model Higgs doublet has a spread over the
entire thick wall, it has complete overlap with the triplet 
Higgs scalar and there is no suppression for the other coupling
$\mu_\xi$. The lepton number violating scale is now given by 
$\l = M_\xi \sim \mu_\xi < M_*$. 

\begin{figure}[thb]
\vskip 0in
\epsfxsize=90mm
\centerline{\epsfbox{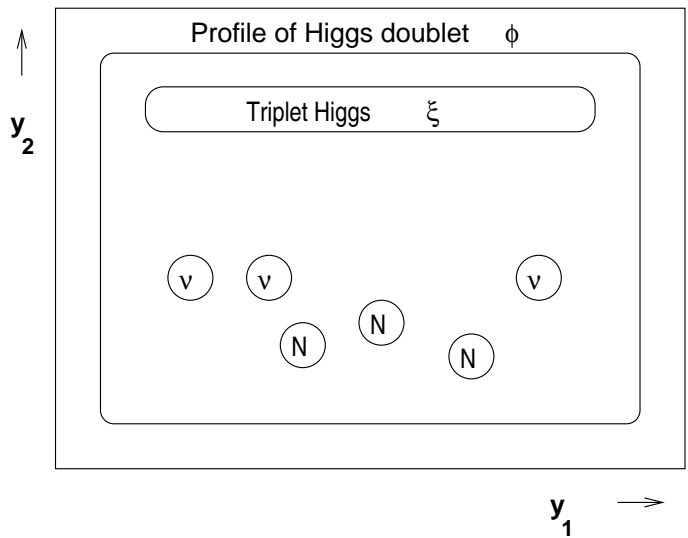}}
\vskip 0in
\caption{
Possible localized positions of neutrinos and the triplet
Higgs with two extra dimensions. 
\label{splfig}}
\end{figure}

So in the thick wall scenario only with a triplet Higgs 
scalar, it is not possible to get a neutrino mass matrix
with the required mixing in the flavor basis. However, an
interesting case may emerge when we include both the triplet 
Higgs scalar and the right-handed neutrinos \cite{lr}. 
We shall also assume two extra dimensions, which is anyway 
required for satisfactory quark mass matrices. Then we
consider the possible localized positions of the left-handed
and right-handed neutrinos and the triplet Higgs, as shown
in figure \ref{splfig}. 

The positions of the left-handed neutrinos are determined
by the charged lepton mass hierarchy and the possible pattern
of the mass matrix. We thus present a scenario of diagonal charged
lepton mass matrix. The Higgs doublet $\phi$ is spread over 
the entire thick brane and all fields have the same overlap with
the Higgs. There will be very little suppression due to the
volume of the Higgs profile $L$ in the extra dimension, but that 
is the same for all the fields and may be absorbed in the definition
of the higher-dimensional coupling constant. 

The triplet Higgs now has equal average distance from all the
three generations of left-handed neutrinos. As a result, the 
contributions to the diagonal elements of the Majorana mass matrix 
of the left-handed neutrinos are equal. On the other hand, in the
basis in which the charged leptons are diagonal, the Majorana mass
matrix generated by the triplet Higgs is also diagonal. So, we have
an exactly degenerate diagonal mass spectrum for the neutrinos 
coming from the triplet Higgs. If we now assume that the lepton
number violating scale is around $\l \sim 10^6 $ GeV, then a
separation between the mean location of the triplet Higgs to the 
mean location of the left-handed neutrinos of $5.35 ~ \mu^{-1}$
would give a neutrino mass matrix
\begin{equation}
m^\nu_{\xi i j} = (0.4 ~ {\rm eV}) \times 
\pmatrix{1&0&0 \cr 0&1& 0 \cr 0&0&1 }   \label{trx1}
\end{equation}
in a basis in which the charged leptons are diagonal. We assumed that
the separation of the left-handed neutrinos is determined by
the charged lepton mass hierarchy as discussed earlier. 

The right-handed neutrino mass matrix will now be most general. 
The mixing angle comes out as required for suitable choice of the
distances in this two-dimensional plane. There is only one 
restriction that a hierarchical neutrino mass spectrum is
only possible numerically, once the left-handed neutrino locations
are determined to get the hierarchy of the charged lepton mass 
matrix. One possible neutrino mass matrix which could emerge
in this scenario is given by
\begin{equation}
m^\nu_{N i j} = (0.025 ~ {\rm eV} )
\pmatrix{0.00004& 0.005& 0.01\cr
0.005& 1.0&  0.9 \cr 0.01& 0.9 & 1.0} . 
\end{equation}
This can explain the solar and atmospheric neutrino anomalies.
This mass matrix predicts an almost maximal mixing with mass
squared difference of $0.002$ eV${}^2$ to explain the atmospheric
neutrinos. A mass squared 
difference of $6 \times 10^{-6}$ eV${}^2$ with a mixing 
angle of $\sin^2 2 \theta \sim 5 \times 10^{-3}$ solves the
solar neutrino problem with the small mixing angle solution. 

A more interesting scenario emerges when the see-saw and the
triplet Higgs scenarios are combined. In this case an almost
degenerate neutrino comes out naturally. It would then be
possible to explain the recent observation of the neutrinoless
double beta decay \cite{ndban} by this hybrid model. In this
case the neutrino mass matrix becomes,
\begin{equation}
m^\nu_{ij} = m^\nu_{\xi i j} + \tilde m^\nu_{N i j} , \label{trx2}
\end{equation}
where $m^\nu_{\xi i j}$ is given by equation (\ref{trx1}). We now
require a different see-saw contribution than the one given for
the hierarchical neutrino mass scheme. In this case it is not
possible to obtain the small mixing angle solution to the solar
neutrino problem. With two extra dimensions, it is possible to
get a mass matrix of the form
\begin{equation}
\tilde m^\nu_{N i j} = (0.0012 ~ {\rm eV} )
\pmatrix{0.1& 1& 1\cr
1& a&  1 \cr 1& 1 & a} ,
\end{equation}
where $a < 0.05$.
The complete neutrino mass matrix given by equation (\ref{trx2})
then gives almost degenerate mass matrix, with all the masses
to be about 0.4 eV. This can explain the neutrinoless double
beta decay and can contribute to the hot component of the dark
matter of the universe. A mass squared difference between $\nu_\mu$
and $\nu_\tau$ of $2.8 \times 10^{-3} ~{\rm eV}^2$  with maximal mixing
solves the atmospheric neutrino anomaly. The mass difference 
between $\nu_e$ and $\nu_\tau$ comes out (for $a = 0$) 
to be $6 \times 10^{-5} 
~{\rm eV}^2$ with mixing angle $\sin^2 ~2 \theta \sim 0.56$, which
can provide the large mixing angle MSW solution to the solar neutrino
problem. 

In summary, we proposed models of neutrino masses in a thick 
wall scenario in which the left-handed and right-handed 
components have split identity. While the model with a right-handed
neutrino can explain the neutrino oscillation experiments, 
the model with triplet Higgs can give only a diagonal neutrino 
mass matrix. Combining the two with two extra dimensions, it 
is possible to obtain an almost degenerate neutrino mass matrix
naturally, which can explain the solar neutrinos, atmospheric
neutrinos, and also contribute to the dark matter of the 
universe and explain the neutrinoless double beta decay.

\vskip .2in
\centerline{\bf Acknowledgement}

\vspace{0.5cm}
\noindent
One of us (U.S.) thanks Max-Planck-Institut f\"ur Kernphysik 
for hospitality.

\end{document}